\documentclass[10pt,pra,twocolumn,floatfix]{revtex4}
\usepackage{amssymb}
\usepackage{amsfonts}
\usepackage{amssymb}
\usepackage{amsmath}
\usepackage[dvips]{graphicx}
\usepackage{color}
\usepackage{appendix}

\usepackage{graphicx}
\usepackage{dcolumn}
\usepackage{bm}

\usepackage{graphicx}

\usepackage{hyperref}

\usepackage{mathrsfs}
\usepackage{isomath}
\usepackage{amsthm}
\usepackage{epstopdf}
\usepackage{txfonts}

\begin{document}

\title{Simultaneous preparation of two optical cat states based on a nondegenerate optical parametric amplifier}

\author{Dongmei Han$^{1}$, Na Wang$^{1}$, Meihong Wang$^{1,2}$, and Xiaolong Su$^{1,2}$}

\email{suxl@sxu.edu.cn}

\affiliation{$^{1}$State Key Laboratory of Quantum Optics and Quantum Optics Devices, Institute of Opto-Electronics, Shanxi University, Taiyuan, 030006, China \\
$^{2}$Collaborative Innovation Center of Extreme Optics, Shanxi University, Taiyuan, Shanxi, 030006, China\\
}

\begin{abstract}
The optical cat state, known as the superposition of coherent states, has broad applications in quantum computation and quantum metrology. Increasing the number of optical cat states is crucial to implement complex quantum information tasks based on them. Here, we prepare two optical cat states simultaneously based on a nondegenerate optical parametric amplifier. By subtracting one photon from each of two squeezed vacuum states, two odd cat states with orthogonal superposition direction in phase space are prepared simultaneously, which have similar fidelity of 60\% and amplitude of 1.2. Compared with the traditional method to generate two odd optical cat states based on two degenerate optical parametric amplifiers, only one nondegenerate optical parametric amplifier is applied in our experiment, which saves half of the quantum resource of nonlinear cavities. The presented results make a step toward preparing the four-component cat state, which has potential applications in fault-tolerant quantum computation.
\end{abstract}

\maketitle

\section{Introduction}

As an important quantum phenomenon, Schr\"{o}dinger cat state~\cite{Schrodinger1935}, which is the superposition of `alive' and `dead' cats, plays an essential role in fundamental physics~\cite{HarocheRMP2013,Arndt2014} and quantum information science~\cite{JeongPRA2002,RalphPRA2003,QEC2022,vanEnkPRA2001,Ulanov2017,Sychev2018,Hacker2019}. It has been experimentally prepared in different systems, such as cavity atomic system~\cite{Leibfried2005,Qin2021}, ion trap~\cite{Monroe1996}, superconducting quantum circuits~\cite{Vlastakis2013,backaction2015,chaosong2019} and optical system~\cite{Ourjoumtsev2006,Neergaard2006,kentaro2007,Takahash2008,Thomas2010,zhang2021,Squeezedcat2022}. The optical cat state, which is defined as the superposition of two coherent states with opposite phases, can be prepared by subtracting photons from a squeezed vacuum state~\cite{Ourjoumtsev2006,Neergaard2006,kentaro2007,Takahash2008,Thomas2010,zhang2021,Squeezedcat2022} and with dissipatively coupled degenerate optical parametric oscillators~\cite{ZYzhou2021,ZYzhou2022}. It is an important quantum resource for quantum communication~\cite{Thomas2010}, quantum computation~\cite{JeongPRA2002,RalphPRA2003,LundPRL2008,Tipsmark2011}, and quantum metrology~\cite{Gilchrist2004,JooPRL2011}.

To realize a quantum algorithm, a series of quantum logic gates need to be implemented, which requires more than one qubit. A large number of physical qubits are the necessary resource in implementing multi-qubit geometric gates~\cite{Chen2022}, realizing nonadiabatic geometric quantum computation~\cite{Kang2022}, and performing quantum error correction algorithms~\cite{Fukui2018}. For example, two odd cat states with orthogonal superposition directions in phase space, $N(| \alpha \rangle - | -\alpha \rangle)$ and $N(| i\alpha \rangle - | -i\alpha \rangle)$, can be encoded as $|0\rangle$ and $|1\rangle$ respectively for the fault-tolerant quantum error correction~\cite{extending2016}, where $N$ is the normalization parameter which is expressed by $1/\sqrt{2(1- e^{-2\left| \alpha \right|^{2}}).}$ Recently, a scheme to generate the optical `four-component cat state', which has potential in fault-tolerant quantum computation, has been proposed \cite{FCCol}. In this scheme, two cat states are coupled on a beam splitter and followed by photon number projection measurement to prepare the four-component cat state. Thus, it is essential to increase the number of cat states and generate cat states with different superposition directions in phase space for scalable and fault-tolerant quantum computation with cat states. 

To generate two cat states with different directions in phase space, a direct scheme is to employ two sets of cat state generation systems. In each system, a degenerate optical parametric amplifier (DOPA) is used to produce a squeezed vacuum state~\cite{Sychev2017}. An approximate odd cat state is heralded if a photon is subtracted from the squeezed vacuum state~\cite{Ourjoumtsev2006,Neergaard2006,kentaro2007,Thomas2010,zhang2021,Squeezedcat2022}. However, such a scheme costs two optical cavities. The nondegenerate optical parametric amplifier (NOPA), on the other hand, enables one to prepare two squeezed states simultaneously from one optical cavity instead of two~\cite{su2012,su2020,hao2021}.

Here, we prepare two independent optical cat states simultaneously based on a NOPA, where the two prepared cat states $N(| \alpha \rangle - | -\alpha \rangle)$ and $N(| i\alpha \rangle - | -i\alpha \rangle)$ have similar properties except the superposition direction in the phase space. At first, two squeezed vacuum states with orthogonal squeezed direction in phase space are produced from a NOPA simultaneously in the experiment. Then, one photon is subtracted from each squeezed vacuum state simultaneously. Finally, the Wigner functions of two photon-subtracted states are reconstructed, which confirms that two cat states with fidelities of  61\% and 60\% and amplitudes of  1.19 and 1.21 are prepared simultaneously in our experiment.

\section{The Principle}

\begin{figure}[h]
\centering
  \includegraphics[width=\linewidth]{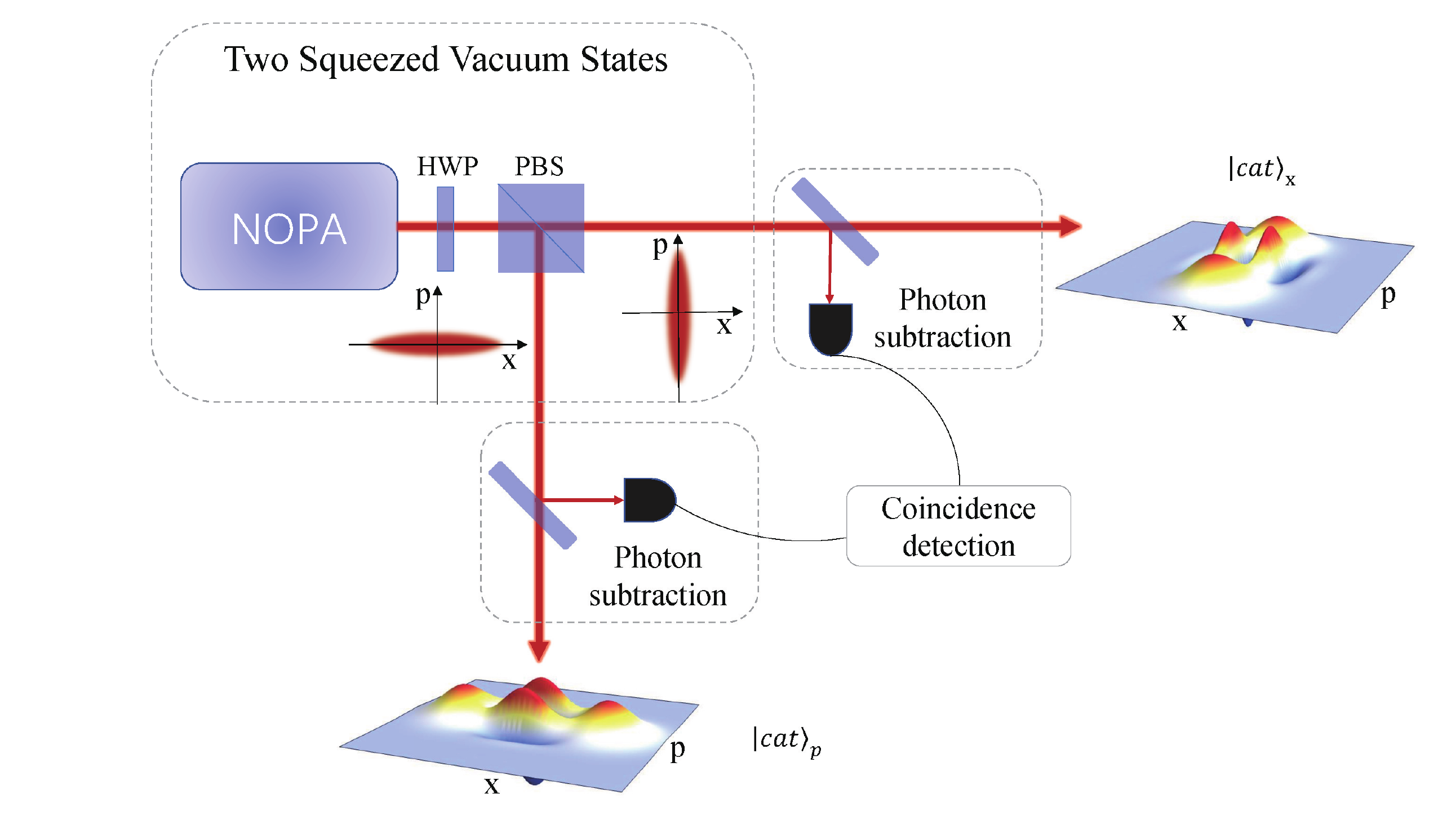}
  \caption{The principle of preparation of two optical cat states simultaneously. HWP: half-wave plate, PBS: polarization beam splitter, NOPA: nondegenerate optical parametric amplifier.}
  \label{fig:boat1}
\end{figure}

The principle of our experiment is shown in Figure 1. When the NOPA is operated at the parametric de-amplification situation and the half-wave plate after the NOPA is set to 22.5 degrees, an amplitude squeezed state and a phase squeezed state are prepared respectively. Then, photon subtraction on each squeezed state is realized with the detection of a small portion of the squeezed mode with a single photon detector. The clicks of coincidence detection from both single photon detectors herald the generation of two optical cat states simultaneously.

The amplitude quadratures $\hat{X}_{1}=(\hat{a}_{1}+\hat{a}^{\dag}_{1})/\sqrt{2}$, $\hat{X}_{2}=(\hat{a}_{2}+\hat{a}^{\dag}_{2})/\sqrt{2}$ and phase quadratures $\hat{P}_{1}=i(\hat{a}^{\dag}_{1}-\hat{a}_{1})/\sqrt{2}$, $\hat{P}_{2}=i(\hat{a}^{\dag}_{2}-\hat{a}_{2})/\sqrt{2}$ of signal mode $\hat{a}_{1}$ and idler mode $\hat{a}_{2}$ from the output of the NOPA are correlated with each other, where the variance of amplitude sum $V[\hat{X}_{1}+\hat{X}_{2}]=e^{-2r}$ and phase difference $V[\hat{P}_{1}-\hat{P}_{2}]=e^{-2r}$ are lower than vacuum noise if the squeezing parameter $r$ is larger than zero ~\cite{NOPA2002,liuyang2022,haijunpr2022,xiaowei2021}. By coupling those two modes on a beam splitter, the coupled modes are expressed by
\begin{equation}
\hat{d}_{+}=\frac{\hat{a}_{1}+\hat{a}_{2}}{\sqrt{2}},
\end{equation}
\begin{equation}
\hat{d}_{-}=\frac{\hat{a}_{1}-\hat{a}_{2}}{\sqrt{2}}.
\end{equation}

The amplitude and phase quadratures of the coupled mode $\hat{c}$ are expressed by $\hat{X}_{c}=(\hat{d}_{+}+\hat{d}^{\dag}_{+})/\sqrt{2}=(\hat{X}_{1}+\hat{X}_{2})/\sqrt{2}$ and $\hat{P}_{c}=i(\hat{d}^{\dag}_{+}-\hat{d}_{+})/\sqrt{2}=(\hat{P}_{1}+\hat{P}_{2})/\sqrt{2}$. The amplitude and phase quadratures of the coupled mode $\hat{d}$ are expressed by $\hat{X}_{d}=(\hat{d}_{-}+\hat{d}^{\dag}_{-})/\sqrt{2}=(\hat{X}_{1}-\hat{X}_{2})/\sqrt{2}$ and $\hat{P}_{d}=i(\hat{d}^{\dag}_{-}-\hat{d}_{-})/\sqrt{2}=(\hat{P}_{1}-\hat{P}_{2})/\sqrt{2}$. The variances of the amplitude and phase quadratures of coupled modes $\hat{c}$ and $\hat{d}$ are given by $V[\hat{X}_{c}]=V[\hat{P}_{d}]=e^{-2r}/2$, which are squeezed in amplitude and phase quadratures, respectively ~\cite{meihong2020}. Thus, one can produce two squeezed vacuum states with the orthogonal squeezed direction simultaneously by using a NOPA. 

The produced amplitude squeezed state $|\psi \rangle_{c}$ and phase squeezed state $|\psi \rangle_{d}$ from the NOPA in the Fock basis are given by \cite{liangguang}
\begin{equation}
|\psi \rangle_{c}=\frac{1}{\sqrt{coshr}}\sum^{\infty }_{m=0}\frac{\sqrt{(2m)!}}{2^{m}m!}(-tanhr)^m|2m\rangle,
\end{equation}
\begin{equation}
|\psi \rangle_{d}=\frac{1}{\sqrt{coshr}}\sum^{\infty }_{m=0}\frac{\sqrt{(2m)!}}{2^{m}m!}(tanhr)^m|2m\rangle,
\end{equation}
where $r$ is the squeezing parameter, and $m$ represents the photon number. After subtracting one photon from each squeezed state, the states we obtained are given by 
\begin{equation}
\hat{a}|\psi \rangle_{c}=coshr^{-\frac{3}{2}}\sum^{\infty }_{m=1}\frac{\sqrt{(2m-1)!}}{2^{m-1}(m-1)!}(-tanhr)^{m-1}|2m-1\rangle,
\end{equation}
\begin{equation}
\hat{a}|\psi \rangle_{d}=coshr^{-\frac{3}{2}}\sum^{\infty }_{m=1}\frac{\sqrt{(2m-1)!}}{2^{m-1}(m-1)!}(tanhr)^{m-1}|2m-1\rangle.
\end{equation}

It has been shown that the state obtained by subtracting a single photon from a squeezed vacuum state approximates an odd cat state when $\alpha \leq 1.2$~\cite{Lund2004}. That is, the states $\hat{a}|\psi \rangle_{c}$ and $\hat{a}|\psi \rangle_{d}$ approximate to the optical cat states $|cat_{x} \rangle$ and $|cat_{p} \rangle$ respectively, which are expressed by

\begin{equation}
|cat_{x} \rangle=\frac{1}{\sqrt{2(1- e^{-2\left| \alpha \right|^{2}})}} (|i\alpha \rangle-|-i\alpha \rangle),     
\end{equation}
\begin{equation}                 
| cat_{p} \rangle=\frac{1} {\sqrt{2(1- e^{-2\left| \alpha \right|^{2}})}}  (|\alpha \rangle-|-\alpha \rangle), 
\end{equation}
respectively. 

\section{The Experiment}

\begin{figure}[h]
\centering
  \includegraphics[width=1.05\linewidth]{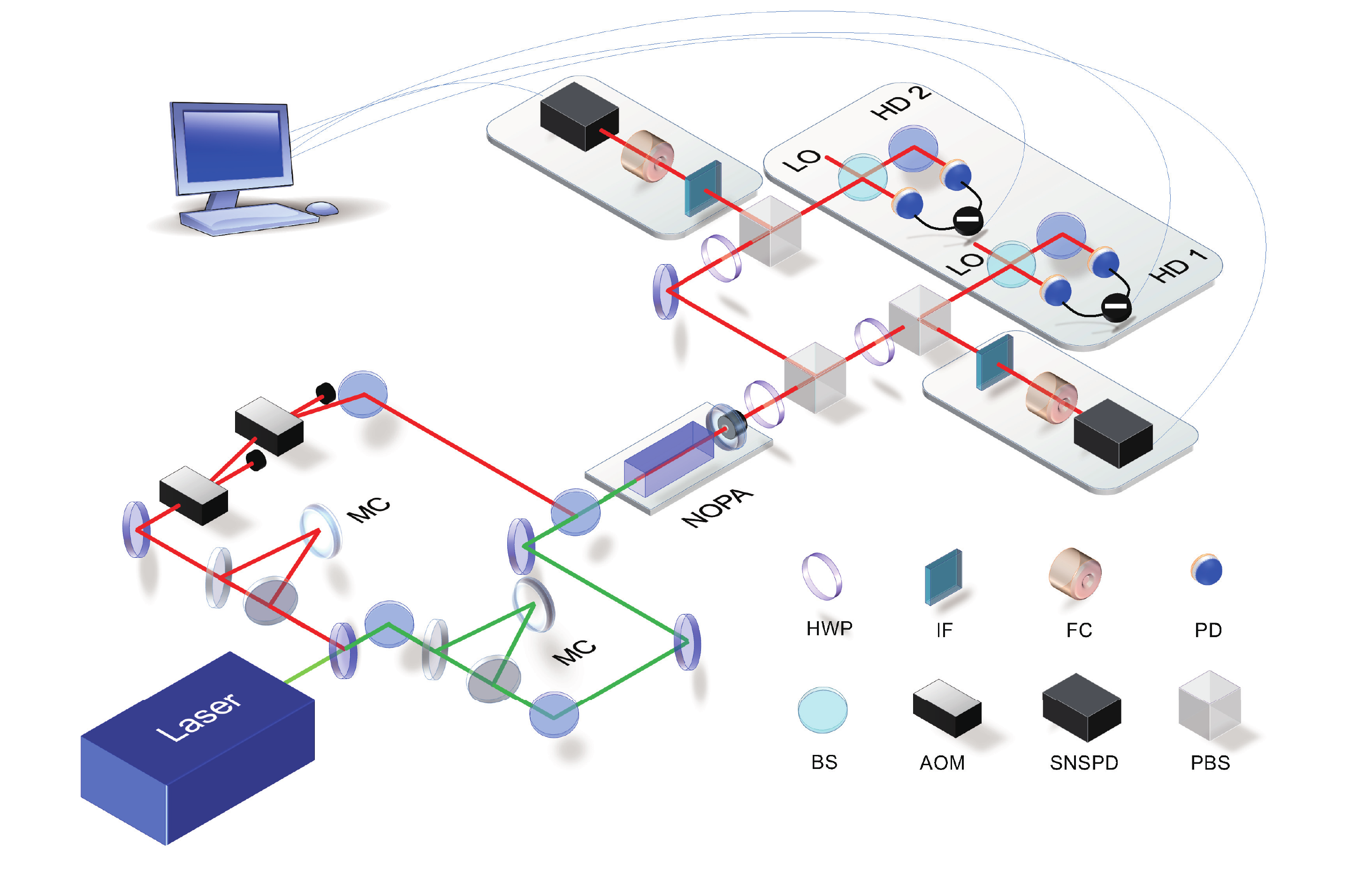}
  \caption{Experimental setup. MC: mode cleaner, NOPA: nondegenerate optical parametric amplifier, EOM: electro-optic modulator, AOM: acousto-optic modulator, PBS: polarization beam splitter, HWP: half-wave plate, SNSPD: superconducting nanowire single-photon detector, HD: homodyne detector, LO: local oscillator, IF: interference filter, FC: filter cavity, PD: photodiode, BS: beam splitter.}
  \label{fig:boat2}
\end{figure}

As shown in Figure 2, a continuous wave intra-cavity frequency-doubled and frequency-stabilized Nd: YAP/LBO (Nd-doped YAIO3 perovskite lithium triborate) laser generates both the 1080 nm and 540 nm laser beams, which are served as the seed beam and the pump beam of the NOPA. Two mode cleaners (MC) are used to filter the spatial and frequency modes of the seed and pump beams respectively. The NOPA consists of a 10 mm $\alpha $-cut KTP (potassium titanyl phosphate) crystal and a concave mirror with a 50 mm radius. The front face of the KTP crystal is coated for the input coupler and the concave mirror serves as the output coupler. The transmissivities of the front face of the KTP crystal at 540 nm and 1080 nm are 40\% and 0.04\%, respectively. The end face of the KTP crystal is antireflection at both 540 nm and 1080 nm. The transmissivities of the output coupler at 540 nm and 1080 nm are 0.5\% and 12.5\%, respectively. The bandwidth of the NOPA is 50 MHz. The triple resonance condition is achieved by adjusting the frequency of the laser and the temperature of the KTP crystal \cite{liuyang2022,haijunpr2022,xiaowei2021}.

The NOPA is operated in the parametric de-amplification situation, where the relative phase difference of the seed and pump beams is locked to $\pi$. When we set the half-wave plate (HWP) at 22.5 degrees, two squeezed vacuum states are generated from two output ports of the polarization beam splitter (PBS) \cite{liuyang2022,haijunpr2022,xiaowei2021}. We inject around 70 mW pump power into the NOPA and generate squeezed vacuum states with squeezing and anti-squeezing levels of around $-$3.2 dB and $4.2$ dB respectively~\cite{han2022ol,han2022prl}.
 
To implement photon subtraction, we use a variable beam splitter which consists of an HWP and a PBS to reflect around 4\% of the squeezed mode toward the superconducting nanowire single-photon detectors (SNSPDs). For the reflected mode of the variable beam splitter, an interference filter with a bandwidth of 0.6 nm and a filter cavity (FC) with a bandwidth of 212 MHz are used to select out the degenerate modes of the NOPA, which have around -39 dB rejection ratio for the nondegenerate modes~\cite{han2022ol,han2022prl}. The FC is locked by using the laser beam reflected by the front mirror of the FC. The transmitted photons of the FC is coupled to the SNSPD to realize photon detection.

\begin{figure}[h]
 \centering
  \includegraphics[width=\linewidth]{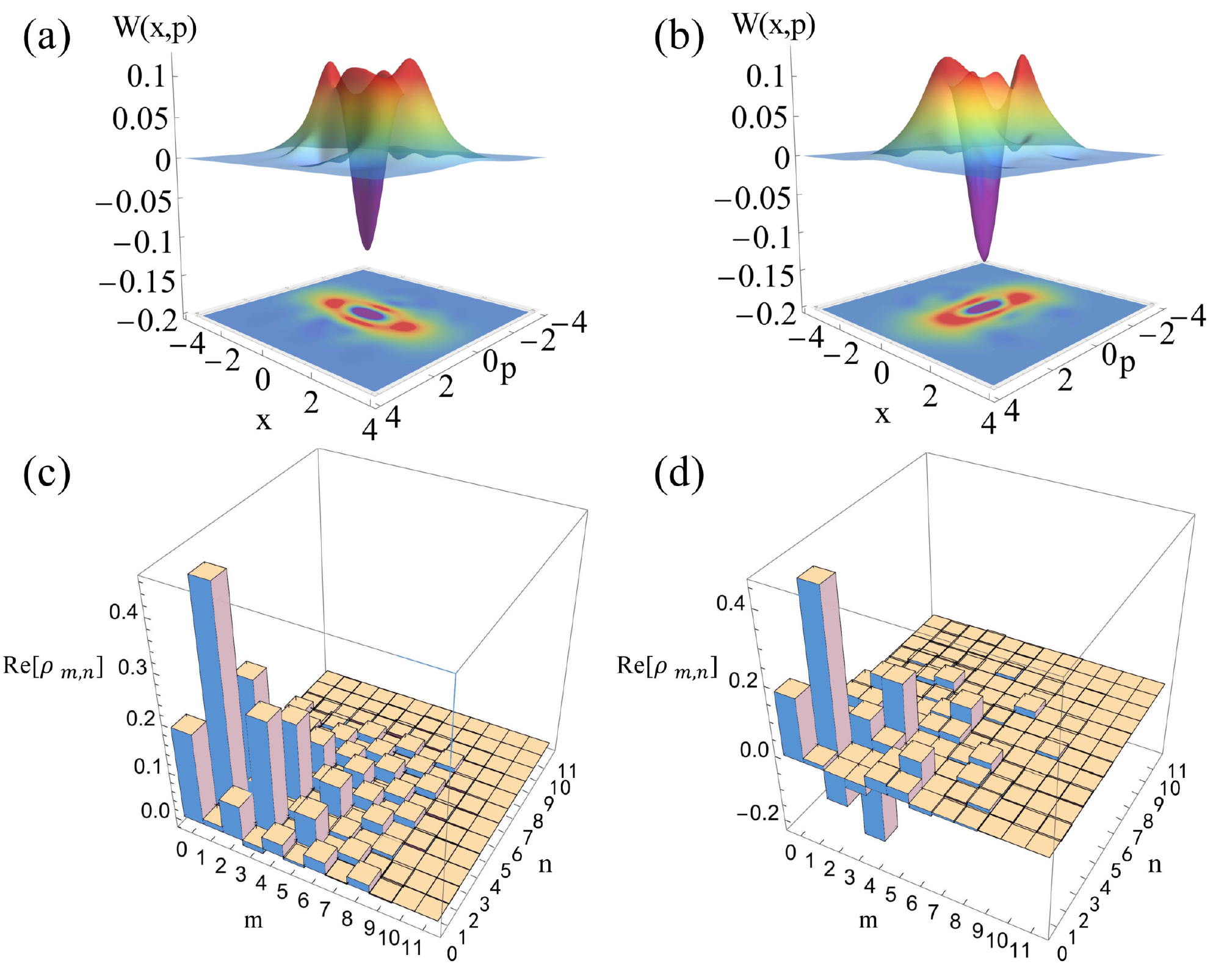}
  \caption{(a, b) Wigner functions and the corresponding contour plots of $|cat_{p} \rangle$ and $|cat_{x} \rangle$, respectively. (c, d) The real parts of the reconstructed density matrix elements of $|cat_{p} \rangle$ and $|cat_{x} \rangle$, respectively. The reconstructed density matrixes are corrected with 81\% efficiency during the reconstruction process which includes the detection efficiency of 90\% of the HD and the transmission efficiency of 90\% of the signal mode.}
  \label{fig:boat3}
\end{figure}

In our experiment, the lock-and-hold technique is applied for NOPA locking~\cite{lockhold2012}, which is performed with the help of two acousto-optic modulators (AOMs). The seed beam of 1080 nm is chopped into a cyclic form with $50$ ms period, which corresponds to each locking and holding period. When AOMs are switched on, the first order of the AOM transmission is injected into the NOPA for cavity locking. Two shutters are closed during the locking period to avoid unwanted trigger events from the SNSPDs. When the AOMs are switched off, the seed beam is chopped off and the NOPA is holding. Two shutters are opened during this period and the tapped-down-converted photons are detected by the SNSPDs. Two squeezed vacuum states are generated and the measurement is performed during the holding period. 

The transmitted modes of the variable beam splitters are transmitted through the optical isolators to isolate the backscattered light from the homodyne detectors (HDs). Then, they are detected by two HDs with a bandwidth of around 80 MHz. The quantum noise limit of the HD is 18 dB above the electrical noise when the local oscillator power is around 18 mW, which means that the clearance of the HD is 96\%. The detection efficiency of HD is around 90\% considering the efficiency caused by the clearance of 96\%, the mode matching efficiency of 98\%, and the quantum efficiency of the photodiode of 98\%. The total transmission efficiencies of two signal modes are around 90\% considering the loss caused by the optical isolators and other optical components.

In order to measure the quantum noise of two signal modes, the AC signals from two HDs are filtered by two 60 MHz low-pass filters respectively and recorded simultaneously by a digital storage oscilloscope (LeCroy WaveRunner) which is triggered by the coincident clicks of two single-photon detectors. The DC signals of two HDs, which represent the interference between signal beams and local oscillators, are also recorded for the phase inference~\cite{likelihood2004}. The sample rate of the oscilloscope is 1GS/s and we record 50000 points for each data file. According to the measured quadrature value $X_{\theta}$ and the corresponding phase $\theta$ of two output states, the Wigner functions of the cat states are constructed by using the Maximum likelihood algorithm with the size of the Fock space up to 12, which is enough for the state with average photon numbers of around 1.2 \cite{likelihood2009}. The generation rate of the two cat states is around 20 Hz when the coincidence window of the single photon trigger events is around 100 ns.

\section{Results}

The reconstructed Wigner functions and corresponding contour plots of the prepared two optical cat states are shown in Figures 3a,b, respectively. It is obvious that cat states $|cat_{p} \rangle$ and $|cat_{x} \rangle$ present the superposition of coherent states along the phase and amplitude quadrature directions respectively. The obtained negativities at the origin of the Wigner functions of two cat states are $-$0.12 and $-$0.14 respectively, which show clearly the non-classical feature of the prepared quantum states. The real parts of the density matrix elements of two prepared cat states are shown in Figures 3c,d, respectively. Compared with cat state $|cat_{p} \rangle$, some elements of the density matrix of the cat state $|cat_{x} \rangle$ are negative, which comes from the fact that the phase of coherent state $|i\alpha \rangle$ is different from $|\alpha \rangle$. In the density matrix, the existence of $|0\rangle$ photon number is caused by the system imperfection in our experiment.

To quantify the quality of the prepared cat state, the fidelity is applied to characterize the overlap between the experimentally prepared state $\rho_{exp}$ and the ideal state $\rho_{ideal}$~\cite{fidelity1994}, which is expressed by $F=Tr[\rho_{exp}.\rho_{ideal}]$. In the case of subtracting one photon from a pure squeezed vacuum state, the fidelity of the generated cat state is decreased with the increase of the squeezing parameter of the squeezed vacuum state, while the amplitude is increased with the increase of the squeezing parameter~\cite{Ourjoumtsev2006,Lund2004}. If the squeezed vacuum state is not pure, the fidelity of the generated cat state is decreased with the decrease of the purity of the squeezed vacuum state~\cite{Laghaout2013}. In our experiment, the purity of the squeezed vacuum state is around 0.9, which limits the fidelity of prepared cat states.

Based on the reconstructed density matrix of the prepared state and the density matrix of the ideal cat state $|cat_{ideal} \rangle$ (shown in Equations 7 and 8), we obtain the fidelities of prepared states. The dependence of the fidelities on the amplitudes of two cat states are shown by blue curves in Figures 4a,b respectively. It is obvious that the maximum fidelities of the prepared states are obtained with amplitudes of 1.19 and 1.21, which means that two cat states $|cat_{p} \rangle$ and $|cat_{x} \rangle$ with fidelities of 61\% and 60\% are experimentally prepared simultaneously.

\begin{figure}[htbp]
\centering
  \includegraphics[width=\linewidth]{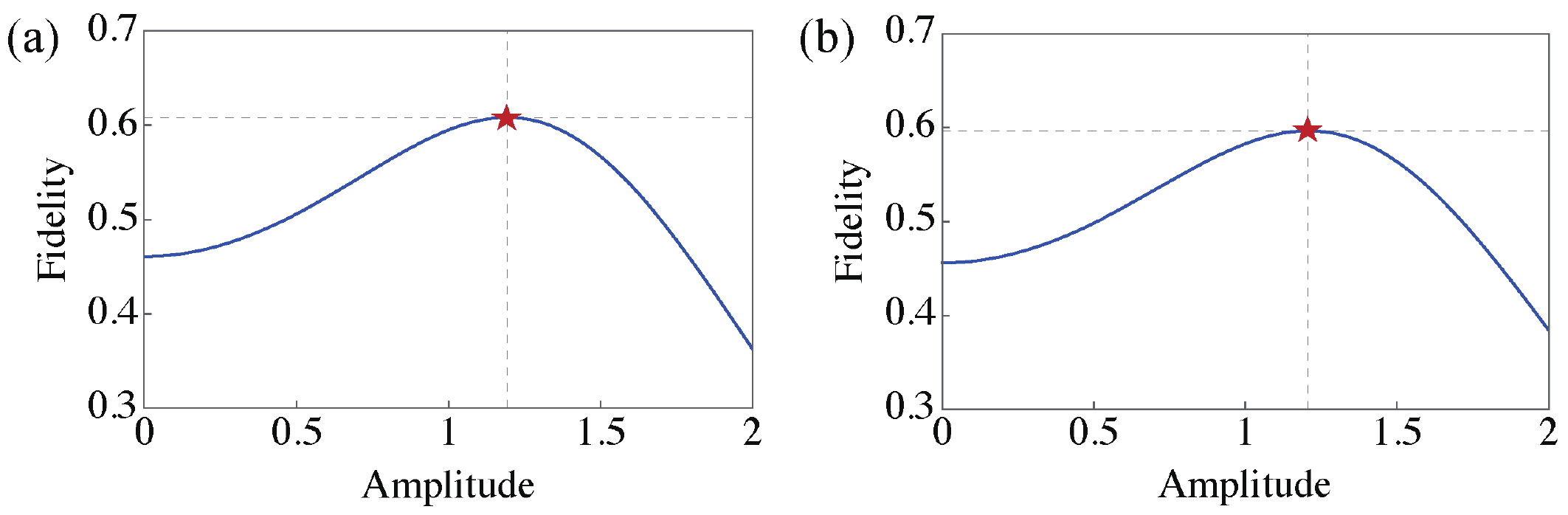}
  \caption{(a, b)Dependence of the fidelity on the amplitude of the prepared cat states $|cat_{p} \rangle$ and $|cat_{x} \rangle$.}
  \label{fig:boat4}
\end{figure}

\section{Discussion and Conclusion}

As an application, our prepared cat states can be used as the input states for the preparation of the optical four-component cat state $| \psi_{FCC} \rangle =  (| \beta \rangle + (-1)^{k} | -\beta \rangle+ (-i)^{k} | i\beta \rangle +i^{k} | -i\beta \rangle)/N_{k}$, where $N_{k}$ ($k=0,1,2,3$) is a normalization factor and $\beta$ is the coherent state amplitude, which has potential applications in fault-tolerant continuous variable quantum computing~\cite{FCCol}. It has been shown that by coupling two single-photon-subtracted states on a 50:50 beam splitter and projecting one output mode of the beam splitter on the Fock state with photon number larger than 2, the other output mode of the beam splitter can approximate the $| \psi_{FCC} \rangle$. Besides this application, the prepared two cat states can also be applied in quantum error correction with cat codes to correct loss errors \cite{extending2016,error2018} and backaction errors \cite{error2018,backaction2015}. 

In summary, we experimentally prepare two optical cat states $|cat_{p} \rangle$ and $|cat_{x} \rangle$ simultaneously by subtracting photons from two squeezed states generated by a NOPA. By reconstructing the Wigner functions of output states, we show that two cat states with fidelities of 61\% and 60\% and the corresponding amplitudes of 1.19 and 1.21 are obtained. Since the advantage of generating two squeezed states from one NOPA is applied in our experiment, we save half of the nonlinear cavity as well as other corresponding instruments compared with the preparation of two cat states based on two DOPAs. Especially, the advantage of our scheme becomes more obvious with the increase of the number of cat states. Besides, we simultaneously prepare two kinds of cat states, i.e. $|cat_{p} \rangle$ and $|cat_{x} \rangle$, with considerable generation rates. Our results demonstrate a new method to prepare cat states with different superposition directions simultaneously and make a step toward fault-tolerant quantum computation with continuous variable error correction code.

\section{ACKNOWLEDGMENTS}

This research was supported by the NSFC (Grants Nos. 11834010 and 62005149), and the Fund for Shanxi \textquotedblleft 1331 Project\textquotedblright\ Key Subjects Construction.

\end{document}